\documentstyle[12pt,psfig]{article}
\begin{document}
\pagestyle{empty}

\begin{center}
{\large \bf A quasiparticle description of the equation of state in SU(3)
     lattice QCD}\\
P. Shukla \\
{\it Nuclear Physics Division,
Bhabha Atomic Research Centre,\\
Trombay, Mumbai 400 085, India}\\
\end{center}

The lattice QCD calculations suggest the appearance of a phase 
transition between the quark gluon dominated high temperature
state and the hadronic state in the low temperature region,
characterized by a large difference in the number of degrees of freedom. 
At temperatures much greater than the critical temperature,
the weak coupling, allows the use of perturbative methods. 
 Near the critical temperature strong nonperturbative effects dominate
the deconfined state. Several models have been proposed
to understand this phenomenon by assuming the appearance of
massive quasiparticles, namely massive quarks and gluons.  
Such a quasiparticle picture has also been invoked in solid
state physics and other fields to study phase transitions in which
a large part of the interaction can be included into the effective masses
of particles which move freely.
  In this work, we analyze recent lattice QCD results
using a phenomenological equation of state with light quarks and 
massive gluons. Such a model would be a preferable
starting point for phenomenological hadronization models
and for a description of the strongly interacting matter near
the phase transition.

 Early lattice data on the pressure and energy density of pure SU(3)
gauge theory (without dynamical quarks) were interpreted with a constant 
gluon mass, $M_g=500$ MeV, and a constant bag constant 
$B^1/4 \simeq 200 MeV$ [1].
Recent SU(3) lattice data were interpreted with an equation of state
where both the gluon mass and bag parameter were temperature dependent [2].
The lattice calculations which include dynamical quarks
based on Wilson fermion formulations
appeared recently, for $N_f=2$ , and $N_f=4$.
Here $N_f$ is the number of flavours.
This data has been used for the extraction of an
equation of state for the QGP containing massive gluons and quarks [3].
Here, the masses and the bag parameter obtained were temperature dependent.

  The lattice data for $N_f=2$ and $N_f=3$ based on staggered 
fermion formulations have also appeared [4],
which we analyze using following equation of state.
 The pressure in the case of a noninteracting gas of 
massive gluons and quarks can be 
written as
\begin{eqnarray}\label{pq}
p_q  = \left( {21 N_f \over 2} f(m_q/T) + 16 \, h(m_g/T) \right)
        {\pi^2 \over 90} T^4- B,
\end{eqnarray}
where $m_q$ and $m_g$ are the masses of quark and gluon,
respectively and $B$ is the bag constant.
The functions $f(m_q/T)$ and $h(m_g/T)$ are given by 
\begin{eqnarray}\label{}
f(m_q/T) &=& (360/7\pi^4) \int_{0}^{\infty}\,du\, u^2
        \ln\left ( 1 + e^{-\sqrt{u^2 + (m_q/T)^2}} \right ) \nonumber \\
{\rm and} \hspace{.1in}
h(m_g/T) &=& -(45/\pi^4) \int_{0}^{\infty}\,du\, u^2
       \ln\left ( 1 - e^{-\sqrt{u^2 + (m_g/T)^2}} \right ) 
\end{eqnarray}

 The figure shows the lattice data [4] for pressure in the case 
of 2 and 3 flavour QGP along with a pure gluon plasma. 
The solid lines are the fit obtained with
Eq.~(\ref{pq}). The fit parameters are given in the table. 
A constant gluon mass and bag constant suffice to give a very good
fit within the errors bars (15 \%) of the data. In principle the 
quark mass should also be made as a parameter, for which
the calculations are under way.
 The QGP produced at RHIC energies is expected to be of three
flavour. Such a simple equation of state presented here 
can be very important for hydrodynamical calculations in the contexts 
of RHIC data.\\
\begin{minipage}{5cm}
\begin{center}
\begin{tabular}{|c|c|c|}
\cline{1-3}
$N_f$  & $m_g$  & $B^{1/4}$ \\
       &  (GeV) & (GeV)     \\
\cline{1-3}
 0 & 0.7 & 0.204  \\
 2 & 1.1 & 0.209  \\
 3 & 1.1 & 0.224  \\
\cline{1-3}
\end{tabular}
\end{center}
\end{minipage}
\begin{minipage}{9cm}
\begin{center}
\psfig{figure=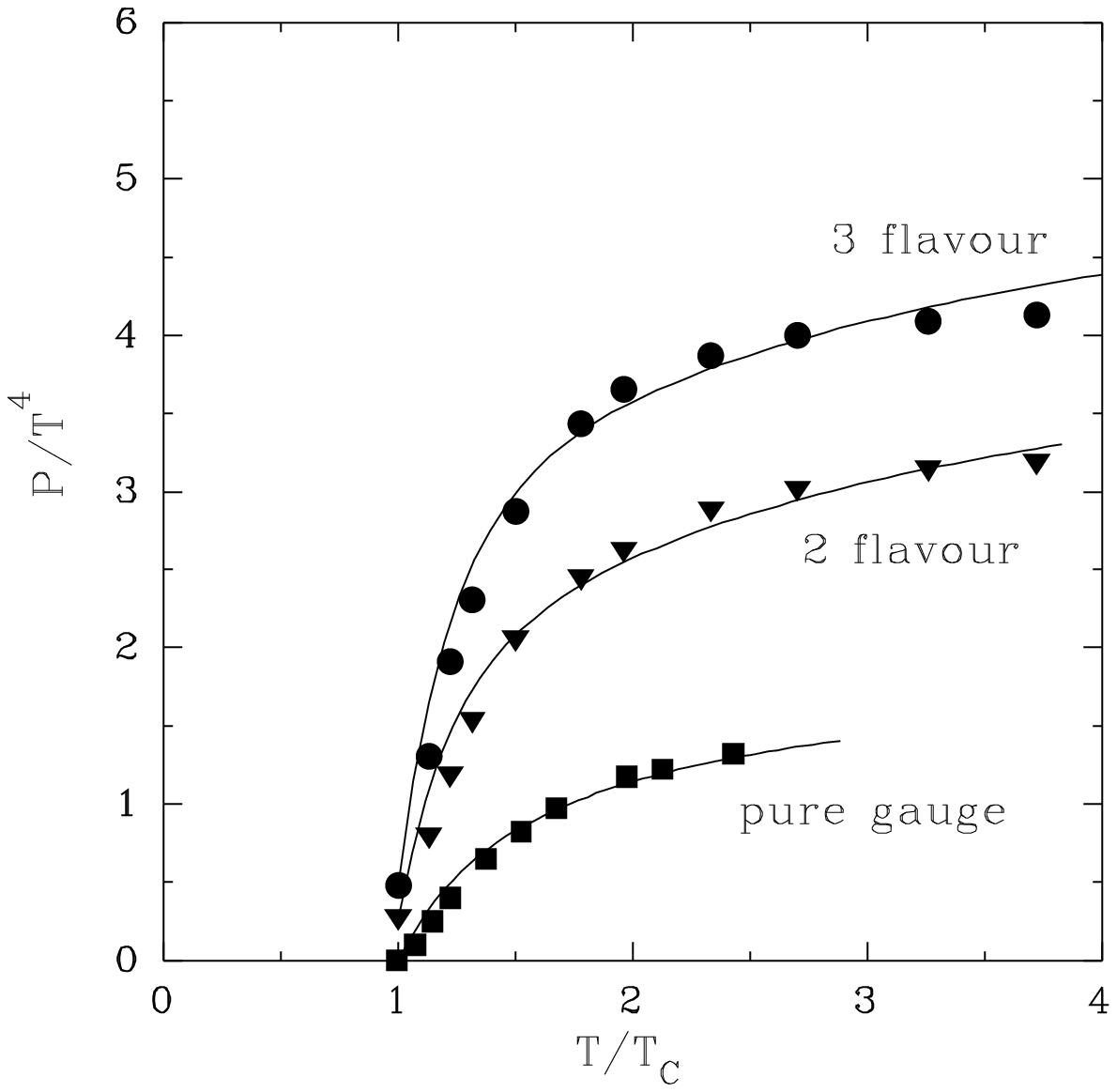,height=5.5cm,width=8.5cm}
\end{center}
\end{minipage}

\noindent
I acknowledge the useful discussions with Dr. A.K. Mohanty.\\

\noindent
[1] T. S. Biro, P.Levai, and B. Muller,
     Phys. Rev. D {\bf 42}, 3078 (1990).

\noindent
[2] A. Peshier, B. Kampfer, O. P. Pavlenko, and G. Soff,
     Phys. Rev. D {\bf 54}, 2399 (1996).

\noindent
[3] P. Levai and U. Heinz,
     Phys. Rev. C {\bf 57}, 1879 (1998).

\noindent
[4] F. Karsch, E. Laermann, and A. Peikert,
    Phys. Lett. {\bf B478}, 447 (2000).

\end{document}